\documentstyle[preprint,aps,floats]{revtex}
\input psfig.tex
\begin{document}
\preprint{PUPT-95-1575, hep-ph/9511238}
\title{Almost-Goldstone Bosons from Extra-Dimensional \\
Gauge Theories}
\author{Neil Turok}
\address{Joseph Henry Laboratory, Princeton University,\\
Princeton NJ, 08544.\\
Email: neil@puhep1.princeton.edu\\
Address after 1/1/96:
DAMTP, Silver St., Cambridge, U.K. }
\date{11/02/95}
\maketitle

\begin{abstract}
A mechanism is presented through which very light scalar degrees of
freedom obeying the nonlinear sigma model equation
can emerge in spontaneously broken gauge theories.
 The mechanism operates 
in extra  dimensional  
theories in which a) there are massless gauge fields present in
the theory prior to compactification, and b) 
the extra
dimensions are inhomogeneous in such a way that
symmetry breaking Higgs fields aquire vevs only
at very localised points on the manifold. These conditions 
are naturally fulfilled in orbifold compactifications 
of string theory. Possible applications include  cosmic texture,
axions and family symmetry.
\\
\end{abstract}

Can the  spontaneous breakdown of a
gauge symmetry produce Goldstone bosons? In normal
circumstances, the Higgs mechanism operates and the Goldstone 
bosons are `eaten' by the longitudinal gauge bosons.
In this Letter, however, I shall show that
in modern approaches to theories
with compact extra dimensions, the answer can be quite different.
Under certain plausible conditions, very light modes
remain which to all intents and purposes are 
the Goldstone bosons associated with the global part of the gauge symmetry.
I call these modes almost-Goldstone bosons (AGB's), because their mass is only 
exponentially small and not precisely
zero.

Goldstone bosons and approximate Goldstone bosons 
are of considerable interest in cosmology, where they provide 
an attractive mechanisms for structure formation in the 
universe. A broken U(1) global symmetry produces cosmic
strings \cite{vs}, 
 a broken nonabelian global symmetry produces cosmic texture \cite{ntx}.
In order for these structure formation mechanisms to work, it is 
essential for the Goldstone bosons 
extremely light, with mass no greater than $\sim 10^{-60} m_{Pl}$. 
Otherwise, the fields settle to their minimum 
and the field ordering process comes to an end, at a time of
order their inverse mass. 
 Similarly, the axion arises
as an approximate Goldstone boson - in order for this mechanism to
solve the strong CP problem, it is essential that any explicit mass 
term be very small, less than $\sim 10^{-38} m_{Pl}$. 
The reason for giving the 
number in Planck units will be made clear below.
Finally, global continuous symmetries are of interest in the context 
of family symmetry (\cite{joyce} and references therein).

Historically, the idea that there could be fundamental global symmetries
has been unpopular in particle theory. 
The gauge principle is 
believed to be a `deeper' idea.
Dynamical theories of the origin of internal symmetries,
such as string theory, naturally produce 
symmetries which are gauged \cite{banks}.
Those  global internal symmetries which are present in the standard model
(related to baryon and lepton number conservation) may 
be explained as being simply `accidents' of the gauge symmetry
and particle content, which could not in general be expected to survive in 
larger unified schemes. And finally, it has been 
argued that quantum gravitational effects
could `spoil' global symmetries by 
introducing terms involving the Planck mass $m_{Pl}$ into
the low energy effective theory which violate global symmetries
in an arbitrary manner \cite{GS,KM}.
These last arguments have to some extent been
countered in \cite{kallosh}.

Without getting into such arguments,
it is still interesting
to ask the following. 
 If we {\it do} accept the proposition that at a fundamental 
level all 
internal 
symmetries are gauged, does it follow that Goldstone bosons of the
kind that are interesting for cosmology are disallowed? I shall show that
the answer is negative.
In particular, if we start with an
extra dimensional 
gauge theory with spontaneous symmetry breaking, under certain 
conditions AGB's with exponentially small masses are produced.
In a compactified field theory, one finds 
$m_{AGB} \propto e^{-ML} $,
 with $M$ a mass scale associated with the symmetry 
breaking Higgs field and $L$ the size of the extra dimensions.
And in  compactifications of string theory, the suppression can be
even stronger - potentially, one has 
$m_{AGB} \propto e^{-T L^2} $ where 
$T$ is the string tension. 

I  shall assume that
there exist gauge bosons {\it before} the theory is
compactified, and that the extra dimensions 
do not possess any special continuous symmetry.
Both assumptions are those usually made in
modern approaches to Kaluza-Klein theory and string theory.

\begin{figure}[htbp]
\centerline{\psfig{file=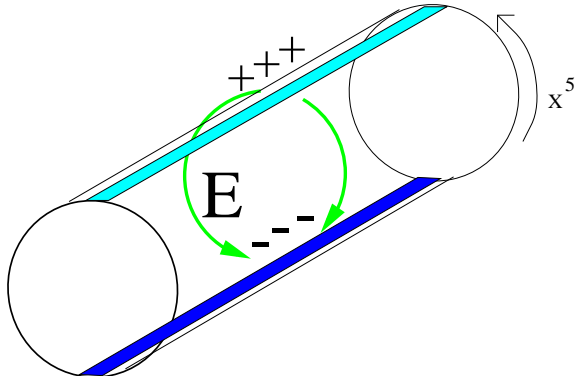,width=3in}}
\caption{ An example of symmetry breaking which is inhomogeneous 
in the extra dimensions, parametrised by $x^5$.
The direction along the tube is an uncompactified dimension.
The Higgs vev is large on the two shaded strips.
A TEM mode can propagate along the tube, with
a positive charge wavepacket travelling along one strip, a negative 
charge wavepacket along the other. E denotes the electric field. 
}
\label{fig:f1}
\end{figure}

As an illustration of the idea, consider 
compactification on
a circle of length $L$.
As mentioned above, I am really interested 
in the case where the extra dimensions {\it do not} possess 
any special symmetries. Therefore I shall ignore any effects 
due to the translational symmetry of the circle. 
Consider  the abelian Higgs model in five dimensions, and assume that 
the Higgs field $\varphi$ gets a vev in the ground state
which is {\it inhomogeneous 
} on the circle (Figure 1).
Let me emphasise that I am just putting this vev in by hand in
this example. The most natural origin for this
inhomogeneity  would be the
inhomogeneity of the compactified dimensions, which
requires more than one dimension, and is therefore harder to
picture.
For example
in orbifold compactifications the curvature has delta function 
singularities at the fixed points of the orbifold.
If $\varphi$ couples to the curvature in 
such a way that its effective mass squared is large and negative at
these points, but positive elsewhere, it will
aquire an inhomogeneous vev
over the extra dimensions, with exponentially small values where 
its mass squared is positive.

The existence of a very light mode follows from the fact
that the two strips  of nonzero Higgs vev are  superconducting
wires, and the configuration shown is a transmission line.
In Lorentz gauge, the equation of motion for the 
gauge field is 
\begin{equation}
\partial^2
A_{\hat{\mu}} = 
 (\Box-\partial_5^2) A_{\hat{\mu}}
= -gv^2(x^5)(\partial_{\hat{\mu}} \theta
+g A_{\hat{\mu}}) \equiv J_{\hat{\mu}}(x)
\label{eq:f1}
\end{equation}
where $\partial^2$ is the full Laplacian, $\Box$ that for the
uncompactified dimensions, and $\partial_5=\partial/\partial x^5$.
Hatted indices run over all dimensions, unhatted indices over
 $0-3$.
The phase $\theta$ is the phase of the Higgs field, and $g$ the
gauge coupling constant.
The Higgs vev $v(x^5)$ is treated as fixed: 
we ignore 
the zero modes corresponding to translation of the wires around the extra
dimensions, a result of the translational symmetry of the 
circle.

In the approximation that the wires are infinitely thin,
$v^2(x^5)=V^2 \sum_i \delta(x^5-x_i)$, 
and we can safely take the current $J_5$ flowing 
around the circle to be zero. We define the Higgs field phase on each wire
to be $\theta^{(i)}$, $i=1,2$, and seek a solution
in which $A_5$ is zero, but $A_\mu= f(x^5) a_\mu(x^\alpha)$.
The form of $f(x^5)$ is simple: between 
the wires it is linear in $x^5$, with slope $\pm f_0$. 
Note that $f(x^5)$ must be periodic in $x^5$.
This follows because we assume that the Higgs field  
has no winding number around the extra dimensions, so 
$\theta$ is single valued. But the currents $J_\mu$ must be single valued,
and therefore so must $A_{\mu}$ be.
Matching the slope discontinuity across 
each wire, 
using (\ref{eq:f1}) we find $2 f_0 a_\mu(x^\alpha)= g V^2 (\partial_\mu
\theta^{(1)} - {1\over 4} gLf_0 a_\mu) $, and similarly with 
$\theta^{(1)}$ replaced by $-\theta^{(2)}$. These equations require that
$\partial_\mu \theta^{(1)} =
-\partial_\mu \theta^{(2)}$.  Thus we find 
\begin{equation}
g A^{(1)}_\mu = - {I \partial_\mu \theta^{(1)} (x^\alpha) \over (1+I)}
= -g A^{(2)}_\mu
\label{eq:f2}
\end{equation}
where $I= {1\over 8} g^2 LV^2$,
and the current flowing on the two wires is 
\begin{equation}
J^{(1)}_\mu = {g V^2 \partial_\mu \theta^{(1)} \over (1+I)} = -J^{(2)}_\mu.
\label{eq:f3}
\end{equation}
The solution to these equations involves one massless 
scalar degree of freedom, $\theta( x^\alpha)$, obeying the
equation for current conservation, $\Box \theta=0$. It 
describes the propagation of a transverse electromagnetic field (TEM) mode
down the transmission line  provided by the extra dimensions. 
Just as in a transmission line, we need at least 
two wires to carry the light mode.
It is straightforward to find all the propagating modes, and to show that
they have masses 
$\sim L^{-2}$, the usual Kaluza-Klein
tower of massive states. 

One simple way of understanding the mechanism is to 
realise that before the symmetry is gauged, the phases
of the Higgs field on the two wires, $\theta^{(1)}$ and $\theta^{(2)}$,
are {\it decoupled} in the limit that $v(x^5)$ vanishes 
in between the wires. Thus
there are actually {\it two} four-dimensional Goldstone modes.
When the gauge field is introduced, it can only `eat' one of these
(the linear combination $\theta^{(1)}+\theta^{(2)}$), leaving the
other massless. The extra dimensions can produce more than one, 
and in principle an infinity of four-dimensional Goldstone modes, and 
thereby  `evade' the Higgs mechanism. 

This discussion generalises to the nonabelian case quite
straightforwardly. The key point is that the only large \cite{correct}
components of the electromagnetic  field tensor are $F_{50}$ 
and $F_{5i}$. Since $A_5$ is zero, neither of these involves any
commutator terms, and the equations of motion are 
the same as in the abelian case, (\ref{eq:f1}), with 
an extra Lie algebra-valued index on the gauge fields and
scalar field current $J_\mu$. 
Apart from adding the extra indices, 
the change in the formulae
(\ref{eq:f2}) and (\ref{eq:f3})
is that $\partial_\mu \theta$ is replaced by 
$-i$ Im($\varphi^\dagger T^a \partial_\mu \varphi)/v^2$, where
$T^a$ are the generators of the gauge group.
In the approximation that all massive modes are set zero, 
conservation of this current is equivalent to
the nonlinear sigma model (NLSM) 
equation. 

This is an important point. In the abelian case,
the compactified theory 
{\it already} has a massless scalar, {\it before} symmetry breaking, 
namely $A_5(x^\alpha)$. 
So it might appear that all we have done is to relabel this mode
in a way that 
makes it look like a Goldstone boson. The nonabelian case
demonstrates that this is not so - the field equation describing 
the low energy excitations  really is the NLSM equation, whereas 
for a dimensionally reduced nonabelian gauge theory $ A^a_5(x^\alpha)$
is instead a massless scalar field in the adjoint representation 
of the gauge group.

We now have all the ingredients needed for the cosmological structure 
formation mechanism. We assume as usual 
that the Higgs vev $<\varphi>$ is zero
at high temperatures,
and a symmetry breaking 
occurs as the universe cools. 
The Higgs field aquires an orientation in internal space which is
 uncorrelated on large (three-dimensional) spatial 
scales.
Then ordering of the fields proceeds, just as in the 
usual scenario \cite{vs,ntx}. Fluctuations of the observed 
amplitude are produced for $V \sim 10^{-3} m_{Pl}$.

We now turn to a more realistic computation, where we include 
the dynamics producing the inhomogeneous Higgs vev.
In this case, we shall find that the mode described above 
gets an exponentially small mass. To be specific, consider the case
where a charged scalar field $\varphi$ has a coupling to the Ricci 
curvature $R$, of
the form $ \xi R \varphi^\dagger \varphi$, with the sign such as to give
$\varphi$ a large negative mass squared at special 
locations on the extra dimensions. Let us make the
approximation
that the scalar curvature $R$ has delta function 
singularities in the extra dimensions. The simplest example would 
be two extra dimensions taking the form of a narrow  tube with
rounded ends, with  strong curvature at either end and none in 
between. Note that whatever the sign of $\xi$  there
is always {\it some} form for the extra dimensions which will
produce the negative mass squared: for example, by adding a  small
handle to each end of the rounded tube just mentioned, one can make the Euler 
number negative so that the mean curvature at each end will
be large and negative \cite{caveat}.

The equation we need to solve for the value of the symmetry breaking field 
in the ground state, $\varphi^\dagger \varphi = v^2(x^5)$  is:
\begin{equation}
-\partial_5^2 v = - \lambda v^3 -m^2 v +\xi \sum_i \delta(x^5-x^i) v(x^i).
\label{eq:f4}
\end{equation}
The symmetry breaking field is assumed to have a positive mass squared 
term and a quartic term in its potential. The solution to
(\ref{eq:f4}) describes a particle rolling up one side of 
the `upside down' potential, approaching the `summit' slowly 
before turning around and descending again. The  delta function terms
reverse the `velocity' $\partial_5v$, sending $v$ rolling back up the hill.
If we assume that the size $L$ of the circle is large, then 
$v(x^5)$ becomes nearly zero in between the delta functions. 
Then the value of $v$ at the delta functions is determined by
energy conservation, ${1\over 2} (\partial_5v)_{max}^2 \approx {1\over 4}
\lambda v_{max}^4 + {1\over 2} m^2 v_{max}^2 $, and 
the matching condition $2 \partial_5 v_{max} = \xi v_{max}$.
Thus we find $v_{max}^2= (\xi^2-4m^2)/2\lambda$. 
A nonzero solution exists if $\xi > 2m$, 
which is just the condition for instability of the configuration $v(x^5)=0
$. In between the delta functions, $v$ gets exponentially small,
$v_{min} \sim e^{-Lm/4} $.
So the symmetry breaking is exponentially small, but nonzero, in between
the wires. 

We would now like to  determine the mass of the lightest mode in 
this more realistic setting. The calculation is most easily 
done in unitary gauge, which is adequate for
considering small fluctuations about the ground state. 
In fact, it is interesting to see exactly 
how the light mode emerges here, because 
in this gauge there are {\it no} degrees of freedom
associated with the phases of the Higgs field. 
Instead we just have a single massive vector boson. Its
equation of motion is 
\begin{equation}
\partial_{\hat{\mu}} F^{\hat{\mu} \hat{\nu}} = - M^2(x^5) A^{\hat{\nu }}
\label{eq:f5}
\end{equation}
where $M^2= g^2 v^2$. This equation implies that $\partial_{\hat{\mu}}(
 M^2(x^5) A^{\hat{\mu}}) =0$, and solving for
 $\partial_{\hat{\mu}} A^{\hat{\mu}}$ one finds 
\begin{equation}
-\Box A_{\hat{\mu}} =  (-\partial_5^2 + M^2(x^5)) A_{\hat{\mu}} 
- \partial_{\hat{\mu}} ({\partial_5 M^2  \over M^2} A_5)
\label{eq:f6}
\end{equation}
For $\hat{\mu}=5$, this only involves $A_5$. 
The right-hand side is a linear operator $\hat{O}$
acting on $A_5$: the eigenvalues of $\hat{O}$ are the squared masses
$M_4^2$ of the four-dimensional fields corresponding to 
$A_5$. 

The operator $\hat{O}$ can be replaced by a hermitian operator $\hat{H}$
if we remove the first order derivative by redefining $A_5= a(x^5)/M(x^5)$.
One finds $\hat{H}=\hat{H}_1
+M^2$, where $\hat{H}_1= 
-\partial_5^2 + M \partial_5^2 (M^{-1}) $. 
Now a minor miracle happens: $\hat{H}_1$ has an eigenvalue which is
{\it exactly } zero, with eigenfunction $a(x^5) = M^{-1}(x^5)$, localised 
in between the wires. We 
obtain a useful upper bound on the 
mass squared of the lightest four-dimensional mode 
by simply using this zero mode of $\hat{H}_1$ as a 
trial function: 
\begin{equation}
M_4^2 < {\int dx^5 (a \hat{H} a) \over \int dx^5 a^2 } = 
{ L \over \int dx^5 M^{-2}(x^5) }.
\label{eq:f7}
\end{equation}
But this means that $M_4^2$ is very small,
because the integral in the denominator has an exponentially
large contribution in between the wires,
where the gauge boson mass 
$M(x^5)$ becomes small. The mass of the lightest
mode is  suppressed as $e^{-Lm/4}$,
 as claimed above.

The mechanism discussed above seems to work even more efficiently
in string theory.
The natural analogs of 
the curvature-coupled Higgs field 
are the so-called twisted modes on orbifolds \cite{dhvw}, which play a 
central role in string phenomenology. Viewed as modes of the string
these excitations have a center of mass which cannot propagate on 
the orbifold,
they 
are literally stuck on the orbifold fixed points. There is 
nothing preventing the scalar fields corresponding to these
twisted modes aquiring a GUT-scale vev,
and the interactions between 
the modes associated with different fixed points are all 
exponentially suppressed.  

Some couplings are suppressed by a Gaussian rather 
than an exponential in the distance between the 
fixed points. 
 This can be  heuristically 
understood as follows. In order for 
 twisted string modes at two different fixed points
to interact, they must undergo
a large quantum fluctuation, in which  their size changes by 
a scale of order the distance $L$ between the fixed points.
Such fluctuations are suppressed by the large 
Euclidean action involved,  the area 
of the string worldsheet times the string tension.
Since the string dynamics is independent
of the tension $T$, dimensional analysis indicates the action 
must be proportional to $L^2$.  The suppression factor is therefore
$\sim e^{-TL^2}$.
The exponent has been calculated in some explicit 
examples by Hamidi and Vafa \cite{hv}, 
and by Dixon, Friedan, Martinec and
Shenker\cite{dfms}. 
So if $L$ is of order ten or
twenty 
string units, the coupling is completely negligible.

The field theory form of the exponential suppression 
may also occur here \cite{dixon}. For example, on a $Z_N$ orbifold $N$
identical twisted strings may combine
at a fixed point into an untwisted state, carrying gauge group indices
for the symmetric product representation, which can then
propagate across to a neighbouring fixed point. In this case
it is natural to conjecture the suppression 
would instead be 
$\sim e^{-ML}$, with $M$ the mass of the intermediate
untwisted charged state, and $L$ the distance between 
fixed points. Again, in string theory the prospect of
obtaining a large exponent is better 
than in field theory  because 
the intermediate state could be forced to be a 
high level number massive string mode. 
As a quasi-realistic example, the $Z_3$ orbifold \cite{dhvw}
has massless gauge fields corresponding to the gauge group
$E_6\times SU(3)$, 
and there are no less than  eighty one $(\bf{1}, \bf{3})$'s
(three at each fixed point).  There is certainly no shortage of
candidate fields for forming
$SU(3)$ global texture in this model.  Indeed one might instead worry that
too many light AGB's might be produced, thus causing conflict with
primordial nucleosynthesis.

What are  the limits to the exponential suppression?
There is a well-known problem associated with making the extra 
dimensions very large, namely that in dimensional reduction, 
the higher dimensional gauge coupling gets large,
$\alpha = \alpha_4 {\cal V}$ where ${\cal V}$ is the volume of the
compactified space, and $\alpha$ as usual is $g^2/4\pi$.
If ${\cal V}$ increases too far, then for fixed four-dimensional coupling
the higher dimensional
theory is in the strong coupling limit, and calculations are impossible.
As a simple example, consider an orbifold, 
with all dimensions small except one. 
The five dimensional coupling would then be potentially problematic.
Let us estimate the largest possible
suppression factor $e^{-ML}$, following the discussion 
of the previous paragraph. Here $L = \alpha_5/\alpha_4$, and
we could consider taking the dimensionless loop expansion parameter 
$(\pi T)^{1\over 2} \alpha_5/(2\pi)$ to be of order unity.
We  use
$\alpha_4 =1/27$, the GUT coupling, and for $M$ take
the mass of a massive string intermediate 
state, $M= \sqrt{8\pi T n}$ with $n $ the level number (for closed strings). 
The suppression factor is then 
$\sim e^{-480 \sqrt{n}}$!

 One should also note that that there 
are other strong constraints on twisted mode couplings
and in general higher powers of 
the exponential suppression factors can be involved. 
This discussion I have presented is of course very preliminary,
one should aim at a general statement about the precise form of the
suppression in an arbitrary term in the full
effective potential. However,
I hope I have succeeded at least in emphasising that
 that 
numerical factors of $2\pi$ are important in exponents,
and that a detailed study would be worthwhile.

The above considerations may be summarised by saying that
compactified extra dimensions can provide a physical reason for the existence 
of several global copies of the original gauge group in the
effective four-dimensional theory. 
Applications of this idea
to realistic orbifold models of cosmic texture, axions and
family symmetry are interesting directions for future work.

I thank M. Bucher, 
A. Farraggi, A. Goldhaber, D. Gross, H. Verlinde, 
F. Wilczek, E. Witten and especially L. Dixon
for very helpful discussions.
This work was partially supported by NSF contract
PHY90-21984, and the David and Lucile Packard Foundation.


\begin{thebibliography}{99}
\bibitem{correct} There is in this case a nonzero contribution to 
$F_{\mu \nu}$, but this is down by $(L/\lambda)^2$ where $\lambda$
is the wavelength of the four-dimensional mode. A more detailed analysis
(N. Turok, to appear) reveals that the nonlinear sigma 
model that arises has an infinite series of higher derivative 
couplings, just in the chiral model description of the 
strong interactions. It is interesting that in the present 
context this series is exactly calculable from the classical field theory.
\bibitem{caveat} Inhomogeneity of the extra dimensions is generic to
solutions in superstring and supergravity theory. However, 
there is at present only an incomplete understanding of what fixes the size
and shape of the extra dimensions.
\end{thebibliography}
\end{document}